\documentclass[aps,twocolumn,showpacs,preprintnumbers,amsmath,amssymb,nofootinbib,superscriptaddress,showkeys]{revtex4}

\usepackage{hyperref}
\usepackage{float}
\usepackage{epsfig}
\usepackage{graphicx}
\usepackage{mathptmx}      
\usepackage{latexsym}
\usepackage{longtable}
\usepackage{dcolumn}

\begin{document}

\title{Bootstrapping the statistical uncertainties of NN scattering data
}

\author{R. Navarro P\'erez}\email{rnavarrop@ugr.es}
\affiliation{Departamento de F\'{\i}sica At\'omica, Molecular y
  Nuclear \\ and Instituto Carlos I de F{\'\i}sica Te\'orica y
Computacional \\ Universidad de Granada, E-18071 Granada, Spain.}
\author{J.E. Amaro}\email{amaro@ugr.es} \affiliation{Departamento de
  F\'{\i}sica At\'omica, Molecular y Nuclear \\ and Instituto Carlos I
  de F{\'\i}sica Te\'orica y Computacional \\ Universidad de Granada,
  E-18071 Granada, Spain.}
  
\author{E. Ruiz
  Arriola}\email{earriola@ugr.es} \affiliation{Departamento de
  F\'{\i}sica At\'omica, Molecular y Nuclear \\ and Instituto Carlos I
  de F{\'\i}sica Te\'orica y Computacional \\ Universidad de Granada,
  E-18071 Granada, Spain.} 

\date{\today}

\begin{abstract} 
\rule{0ex}{3ex} We use the Monte Carlo bootstrap as a method to
simulate pp and np scattering data below pion production threshold
from an initial set of over 6700 experimental mutually $3\sigma$
consistent data. We compare the results of the bootstrap, with 1020
statistically generated samples of the full database, with the
standard covariance matrix method of error propagation.  No
significant differences in scattering observables and phase shifts are
found. This suggests alternative strategies for propagating errors of
nuclear forces in nuclear structure calculations.
\end{abstract}

\maketitle

\section{Introduction}

The modern era of high quality NN interactions started when the long
term studies of the Nijmegen group culminated in a succesfull least
squares fit with a statistically significant $\chi^2/\nu \sim
1$~\cite{Stoks:1993tb} after implementation of many small but crucial
effects and $3\sigma$ inconsistent data were excluded.  Since then,
subsequent analyses have been carried
out~\cite{Stoks:1994wp,Wiringa:1994wb,Machleidt:2000ge,Gross:2008ps,NavarroPerez:2012qf,Ekstrom:2013kea,Perez:2013mwa,Perez:2013jpa,Perez:2013oba}
having $\chi^2/\nu \sim 1$ and with the purpose of being used in {\it
  ab initio} Nuclear Structure calculations. As is well
known~\cite{evans2004probability} any least-squares fit, corresponds
to $\chi^2$ minimization
\begin{eqnarray}
\min_{\bf p} \chi^2({\bf p}) = 
\min_{\bf p} 
\sum_{i=1}^{N} 
\left( 
\frac{O_i^{\rm exp}-O_i({\bf p})}{\Delta O_i^{\rm exp}}
\right)^2 
\equiv 
\chi^2 ({\bf p}_0) \, ,  
\label{eq:chi2}
\end{eqnarray}
where $O_i^{\rm exp}$ is a fitted observable, $\Delta O_i^{\rm exp}$
the corresponding statistical error bar and $O_i({\bf p})$ the
theoretical model depending on fitting parameters ${\bf p}={p_1,
  \dots, p_P}$. The procedure assumes that the statistical
uncertainties of the fitted data can be modeled by a probability
distribution; namely independent normally distributed data $N(O_i^{\rm
  exp},\Delta O_i^{\rm exp})$ an assumption based on counting a
large number of events in NN scattering experiments. The assumption of
a finite number of normally distributed data is an indispensable
prerequisite for both a meaningful uncertainty estimates from a
phenomenological fit and any subsequent and reliable error
propagation.  Fortunately, normality can be checked {\it a posteriori}
in probabilistic terms and within a given confidence level by
application of a variety of statistical tests, which naturally become
more stringent with the number of data.  Of course, individually
checking the probability distribution of over 6700 data points,
involving over 300 experiments, some dating back more than 60 years,
is rather impractical. However, if a model fitted to the data is
flexible enough to accurately reproduce them, the normality of the
experimental data implies that discrepancies between theory and
experiment, known as {\it residuals}, must follow a standard normal
distribution, i.e.
\begin{equation}
 R_i = \frac{O_i^{\rm exp} - O_i({\bf p}_0)}{\Delta O_i^{\rm exp}} \sim N(0,1),
 \label{eq:normalresiduals}
\end{equation}
Once a fit has been made, testing Eq. (\ref{eq:normalresiduals}) is
straightforward. Despite its simplicity, normality testing has not
been a common practice in nuclear interactions fitting (an early
discussion on normality was however conducted in
Ref.~\cite{Bergervoet:1988zz,Stoks:1992ja}).

In a recent publication~\cite{Perez:2013jpa} we presented a new
phenomenological Nucleon-Nucleon (NN) potential that accurately
describes 6713 scattering data from 1950 to 2013 upgrading much of the
previous works and increasing the statistics. This was done with an
eye put on the determination of the uncertainties in the fitted NN
interaction itself and their consequences in Nuclear Physics, for
which little is still known (see however
\cite{NavarroPerez:2012vr,Perez:2012kt,NavarroPerez:2012qf}) and
statistical methods offer the most natural framework.  On a more
general level, a growing concern on the statistical analysis of
nuclear theory and its predictive power has been initiated~(see
e.g. \cite{dudek2013predictive,Dobaczewski:2014jga} for general and
instructive overviews and references therein). We have applied some of
the well known normality tests to three of our NN potentials,
including the delta-shell potential with one pion exchange
(DS-OPE)~\cite{ Perez:2013mwa,Perez:2013jpa}, (chiral) two pion
exchange (DS-$\chi$TPE)~\cite{Perez:2013oba} and a gaussian potential
with OPE~\cite{Perez:2014yla} and found the normality condition to
hold in all of them~\cite{Perez:2014yla}. The lack of normality would
clearly signal an inconsistency in the fitting analysis and might be
used as a guide to unveil systematic errors both in the data as well
as in the model.  It is thus foreseeable that normality tests will be
regarded as an important ingredient in the design of NN interactions
statistically inferred from scattering data (see
e.g. \cite{Ekstrom:2014dxa} for a posteriori analysis of
\cite{Ekstrom:2013kea}).

In our previous works the covariance matrix method was used to
propagate errors. In the present note we discuss the robustness of our
results using Monte Carlo techniques and the bootstrap
method~\cite{efron1982jackknife}. While these methods have succesfully
been exploited (see e.g.~\cite{Nieves:1999zb,Nieves:1999bx} for
related studies within $\pi\pi$ scattering error analyses) to our
knowledge they have never been implemented within the context of the
NN force, so our presentation will be intentionally pedagogical. We
also outline interesting consequences regarding strategies for error
propagation in nuclear physics.

\section{Covariance matrix method}

In the standard covariance method one starts with a least squares
fit~Eq.~(\ref{eq:chi2}). Once the condition of normality,
Eq.~(\ref{eq:normalresiduals}), has been checked~\cite{Perez:2014yla}
and {\it assuming} normality of errors in the fitting parameters we
are in position to propagate the statistical uncertainties into the
potential parameters and any calculation that takes this potential
parameters as an input. The error matrix ${\cal E}_{ij}$ of the
potential parameters $\{p_1,p_2,\ldots,p_P\}$ can be calculated by
inverting the Hessian matrix
\begin{equation}
\label{eq:HessianMat}
H_{ij} = \frac{\partial^2 \chi^2}{\partial p_i \partial
  p_j}\Big|_{{\bf p}_0}\equiv ({\cal E}^{-1})_{ij} \,
\end{equation}
which can be used to obtain confidence intervals for the parameters
and correlations among them. Any quantity that can be calculated
as a function of the potential parameters $F(p_1,p_2,\ldots,p_P)$ can
be provided with an statistical error bar $\Delta F$ with the
customary expression
\begin{equation}
\label{eq:CovarianceErrors}
(\Delta F)^2 = \sum_{ij} \frac{\partial F}{\partial p_i}
\frac{\partial F}{\partial p_j} {\cal E}_{ij}.
\end{equation}
A good approximation to Eq.(\ref{eq:HessianMat}) can be found
in~\cite{Perez:2013jpa} which has been used with
Eq.(\ref{eq:CovarianceErrors}) to estimate statistical uncertainties
of phase-shifts, scattering amplitudes, deuteron properties, form
factors, matrix elements and skyrme
parameters~\cite{Perez:2013mwa,Perez:2013jpa,Perez:2013oba,Amaro:2013zka,Perez:2014yla,Perez:2014kpa}. However
the derivatives in Eq.(\ref{eq:CovarianceErrors}), depending on the
functional form of $F$, may be hard to calculate analytically. If one
contemplates numerical evaluation this requires a repeated evaluation
of the function $F$ at several values of the fitting parameters, which
for a large number of parameters ( typically 30-40
\cite{Perez:2013mwa,Perez:2013jpa,Perez:2013oba} ) may also be a
costly procedure~\footnote{It can also be an innacurate procedure
  since the corresponding finite differences step $h$ must be smaller
  than the statistical $\Delta p$ which are usually quite small. For
  instance, the evaluation of the Hessian numerically for our fits
  \cite{Perez:2013mwa,Perez:2013jpa,Perez:2013oba} requires to compute
  crossed derivatives, which turned out to be highly unstable for large
  number of parameters.  This is why we prefered to compute the
  derivatives analytically and use in passing the highly efficient
  Levenberg-Marquardt minimization algorithm where a stable (definite
  positive) approximation to the Hessian is
  exploited~\cite{press2007numerical}.}.

The calculation of derivatives can be avoided by drawing random
numbers following a multivariate normal distribution determined by the
covariance matrix ${\cal E}$, 
\begin{equation}
\label{eq:multinormal}
 P(p_1,p_2,\ldots,p_P) = \frac{1}{\sqrt{(2 \pi)^P \det {\cal E}}}
 e^{-\frac{1}{2}({\bf p}- {\bf p}_0)^T {\cal E}^{-1} ({\bf p}- {\bf
 p}_0)},
\end{equation}
This generates a family of potential parameters and calculate $F$
with each potential. This Monte Carlo method directly propagates
uncertainties, however a multivariate normal probability distribution
to all the parameters is assumed which may not always be the case for
the \emph{true} distribution of parameters.

\section{The bootstrap method}

The Bootstrap is a Monte Carlo technique that allows to find the most
likely parameters probability distribution and propagate statistical
uncertainties and correlations into any function
$F$~\cite{efron1982jackknife} (see also \cite{press2007numerical}). In
our case the deviations between the theoretical model and the
experimental data are normal statistical fluctuations the procedure
corresponds to generate replicas of the observed data which are meant
to simulate a fictitious experiment.  Thus, for every experimental
data point $O_i^{\rm exp}$ with uncertainty $\Delta O_i^{\rm exp}$ one
generates $M$ ``synthetic'' random points $O_{i,1}^{\rm
  synth},O_{i,2}^{\rm synth},\ldots,O_{i,M}^{\rm synth}$ distributed
as $N(O_i^{\rm exp},\Delta O_i^{\rm exp})$, i.e.
\begin{equation}
 O_{i,\alpha}^{\rm synth} = O_i^{\rm exp} + \xi_{i,\alpha} {\Delta O_i^{\rm exp}}
 \label{eq:synth-res}
\end{equation}
where $ \xi_{i,\alpha} \sim N(0,1)$ are standard normal and
independent variables,$\langle \xi_{i,\alpha} \rangle =0 $ and
$\langle \xi_{i,\alpha} \xi_{j,\beta} \rangle = \delta_{ij}
\delta_{\alpha\beta} $. This will generate $M$ independent databases
with the same number of data as the original one. Each synthetic
database will represent a snapshot of the random fluctuations inherent
to the experimental processes.  A least squares fit to every generated
database, $O_{i,\alpha}^{\rm synth}$ ($\alpha=1, \dots, M$), featuring
a maximum likelihood estimate can be made and a family of parameters
${p_{1,\alpha}, \dots , p_{P,\alpha}}$ will be obtained as 
\begin{eqnarray}
\min_{\bf p} \chi_\alpha^2({\bf p}) = 
\min_{\bf p} 
\sum_{i=1}^{N} 
\left( 
\frac{O_{i,\alpha}^{\rm synth}-O_i({\bf p})}{\Delta O_i^{\rm exp}}
\right)^2 
\equiv 
\chi^2 ({\bf p}_\alpha) \, . 
\end{eqnarray}
Then, the most likely theory parameters are ${\bf p}_\alpha$. The
corresponding joined or marginal probability distributions can be
obtained by binning the outcoming parameter samples. This allows to
compute any function of the theoretical model parameters $F({\bf p})$
at a set of points $F_\alpha \equiv F({\bf p}_\alpha)$. Thus, the mean
and variance can be computed for large $M$ as usual,
\begin{eqnarray}
E( F)  &=& \frac1{M}\sum_{\alpha=1}^M F({\bf p}_\alpha) \, ,  \\
{\rm Var}(F) &=& \frac{M}{M-1} E \left[( F-E(F))^2 \right] \, , 
\label{eq:variance}
\end{eqnarray}
The correlation coefficient of two different observables is
\begin{eqnarray}
{\cal C} (F,G) = \frac{ E\left[ (F-E(F)) (G-E(G)) \right]}
{\sqrt{E\left[(F-E(F))^2 \right]}\sqrt{E\left[(G-E(G))^2 \right]}} \, , 
\end{eqnarray}
so that ${\cal C} (p_i, p_j)={\cal C}_{ij}= {\cal E}_{ij}/( {\cal
  E}_{ii} {\cal E}_{jj})^\frac12 $ is the correlation matrix. For
asymmetric or skewed distributions, it may be better to define the
$1\sigma$ asymmetric coverage by excluding $16\%$ of the upper and
lower values of the distribution instead of the variance definition,
Eq.~(\ref{eq:variance}).  At any rate we always check this possibility
before errors are quoted.

While the bootstrap method requires to perform $M$ repeated fits,  
it is a competitive alternative to determine errors and
correlations when the covariance matrix itself is not directly
available nor used in the minimization
method~\cite{Kortelainen:2010hv}. Again, we stress that this method to
generate snapshots of the statistical fluctuations is justified since
the condition of Eq.(\ref{eq:normalresiduals}) has been checked to a
significant confidence level.

\begin{figure*}
\centering
\includegraphics[width=\linewidth]{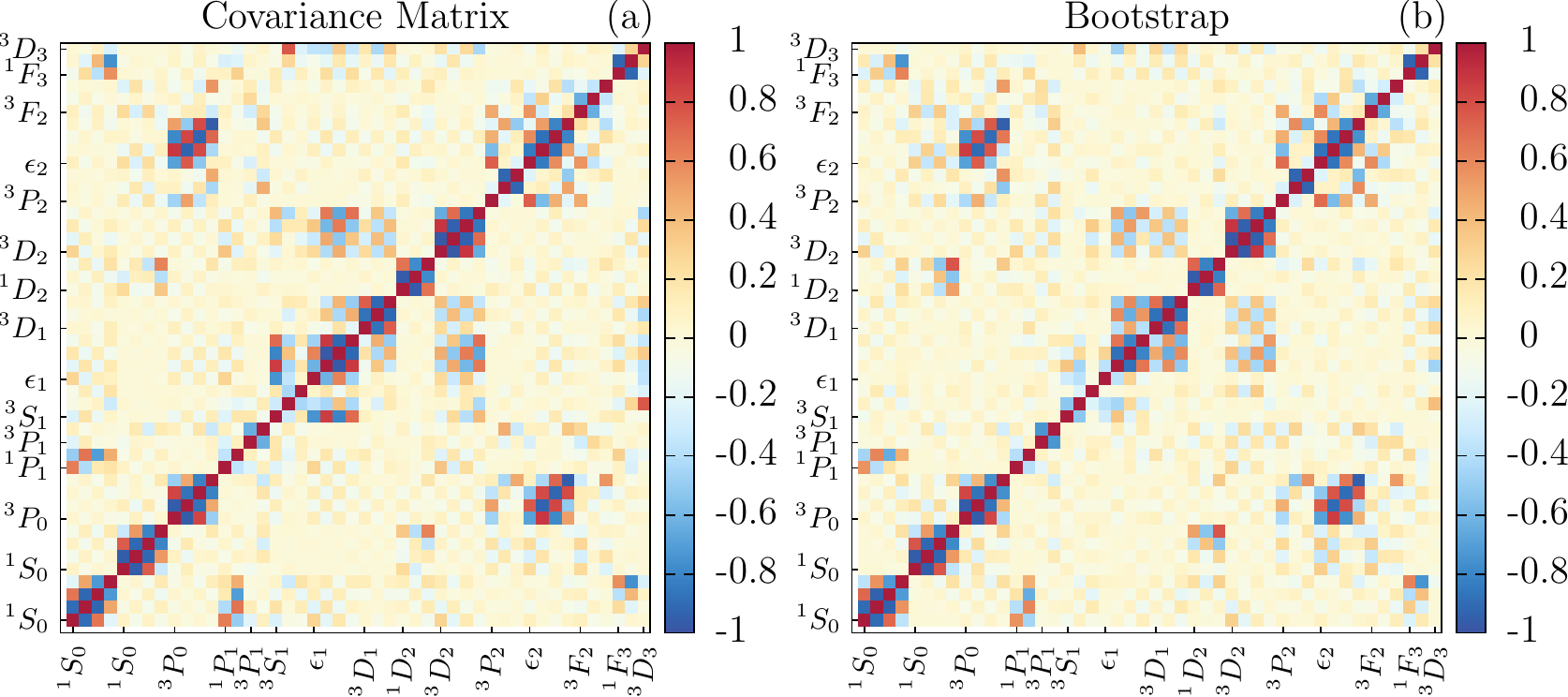}
\caption{(Color online) Correlation matrix ${\cal C}_{ij}$ for the
  DS-OPE potential parameters $(\lambda_i)_{l,l'}^{JS}$ in the partial
  wave basis~\cite{Perez:2013mwa}. The points $r_i = \Delta r (i+1)$
  are grouped within every partial wave. We show the results obtained
  with the covariance matrix (left panel) and the Monte Carlo
  bootstrap simulation of experimental data (right panel). We grade
  gradually from $100\%$ correlation, ${\cal C}_{ij}=1$ (red), $0\%$
  correlation, ${\cal C}_{ij}=0$ (yellow) and $100\%$
  anti-correlation, ${\cal C}_{ij}=-1$ (blue).}
\label{fig:Correlation}       
\end{figure*}

\section{Numerical results}

We apply the different methods to the $3\sigma$ self-consistent
database presented in Ref.~\cite{Perez:2013jpa} where
$\chi^2/\nu=1.04$. The potential used for this analysis has the form  
\begin{eqnarray}
   V(\vec r) = V_{\rm short} (r) \theta(r_c-r)+ V_{\rm long} (r) \theta(r-r_c)\, . 
\label{eq:potential}
\end{eqnarray}
The long range piece $V_{\rm long}(\vec r)$ contains a
Charge-Dependent (CD) One pion exchange (OPE) with a fixed 
$f^2=0.075$~\cite{Stoks:1992ja}) and
electromagnetic (EM) corrections which are kept fixed throughout the
fitting process. The short component was inspired by
Avil\'es~\cite{Aviles:1973ee} (see also
\cite{Entem:2007jg,NavarroPerez:2011fm}) and reads
\begin{eqnarray}
   V_{\rm short}(\vec r) = \sum_{n=1}^{21} \hat O_n \left[\sum_{i=1}^N V_{i,n} \delta(r-r_i)\right] \, , 
\label{eq:potential-short}
\end{eqnarray}
where $ \hat O_n$ are the set of operators in the extended AV18
basis~\cite{Wiringa:1994wb,NavarroPerez:2012vr,Perez:2012kt,Amaro:2013zka},
$V_{i,n}$ are fitting parameters and $r_i=\Delta r (i+1)$ with $\Delta
r=0.6 {\rm fm}$. The fit is carried out more effectively in terms of
some low and independent partial waves contributions to the potential
$(\lambda_{i,\alpha})_{l,l'}^{JS}$ from which all other higher partial
waves are consistently deduced (see
Ref.~\cite{Perez:2013mwa,Perez:2013jpa}).  The delta-shell potential
reduces the computational effort enormously, so a large number of fits
can easily be undertaken.

For the bootstrap analysis we took $M=1020$ samples of the $N=6713$
data and refitted the parameters of the DS-OPE potential (denoted by
$(\lambda_{i,\alpha})_{l,l'}^{JS}$) which was used to determine the
database. This generates $M$ independent sets of most likely
parameters to each synthetic database $O_{i,\alpha}$, $\alpha=1,
\dots, M$. From there any function of the fitted parameters and the
inherent correlations can be determined.

\begin{figure*}
\centering
\includegraphics[width=\linewidth]{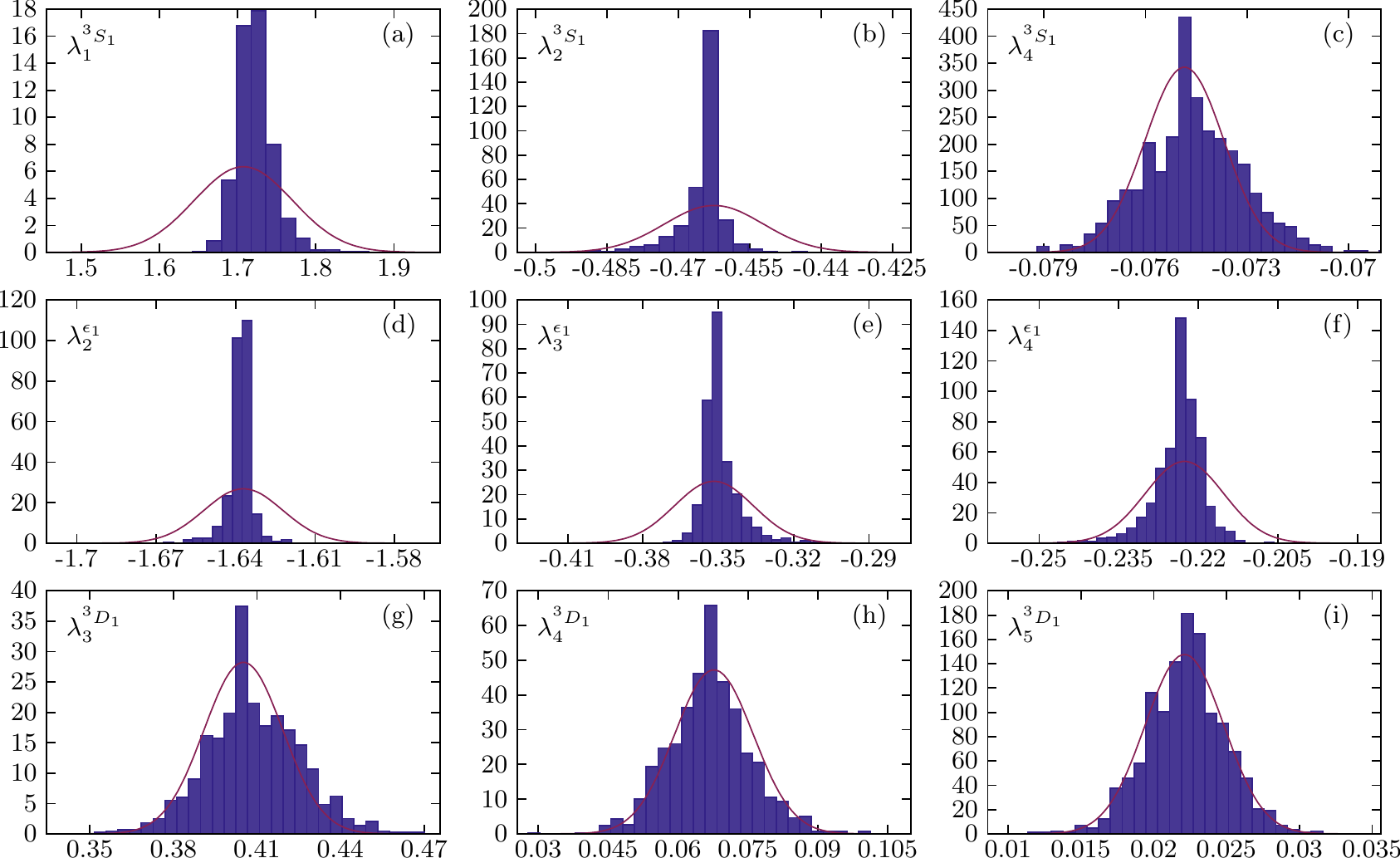}
\caption{(Color online) $^3S_1$-$^3D_1$ coupled channel Delta Shell
  parameters distribution. The parameters are $^3S_1$ partial wave
  (upper row), $\epsilon_1$ mixing angle (middle row) and $^3D_1$
  partial wave (lower row). The blue bars give the normalized
  histogram from the $1020$ fits to the Monte Carlo generated
  databases. The red line is the normal distribution given to each
  parameter by the covariance matrix method. }
\label{fig:J1Distributions}       
\end{figure*}

In Figure \ref{fig:Correlation} we show the correlation matrix of the
DS-OPE potential parameters obtained with the standard covariance
matrix method and the Bootstrap method.  It is not obvious, though
most wellcome, that both covariance and bootstrap methods give fairly
similar results, although small correlations are overestimated by the
covariance matrix. The main difference between both methods is
in the $^3S_1$-$^3D_1$ coupled channel. The Monte Carlo bootstrap
simulation results in very small correlations between the $^3S_1$ and
$\epsilon_1$ partial wave parameters and stronger correlations between
$\epsilon_1$ and $^3D_1$; in contrast the covariance matrix method
gives opposite results. These discrepancies could be related to the
fitting of the deuteron binding energy where the approximation used
for the Hessian matrix might be outside of its range of validity. In
fact, the Monte Carlo generated $^3S_1$, $\epsilon_1$ and $^3D_1$
parameters show large asymmetries as can clearly be seen in Figure
\ref{fig:J1Distributions}.

\begin{figure*}
\centering
\includegraphics[width=\linewidth]{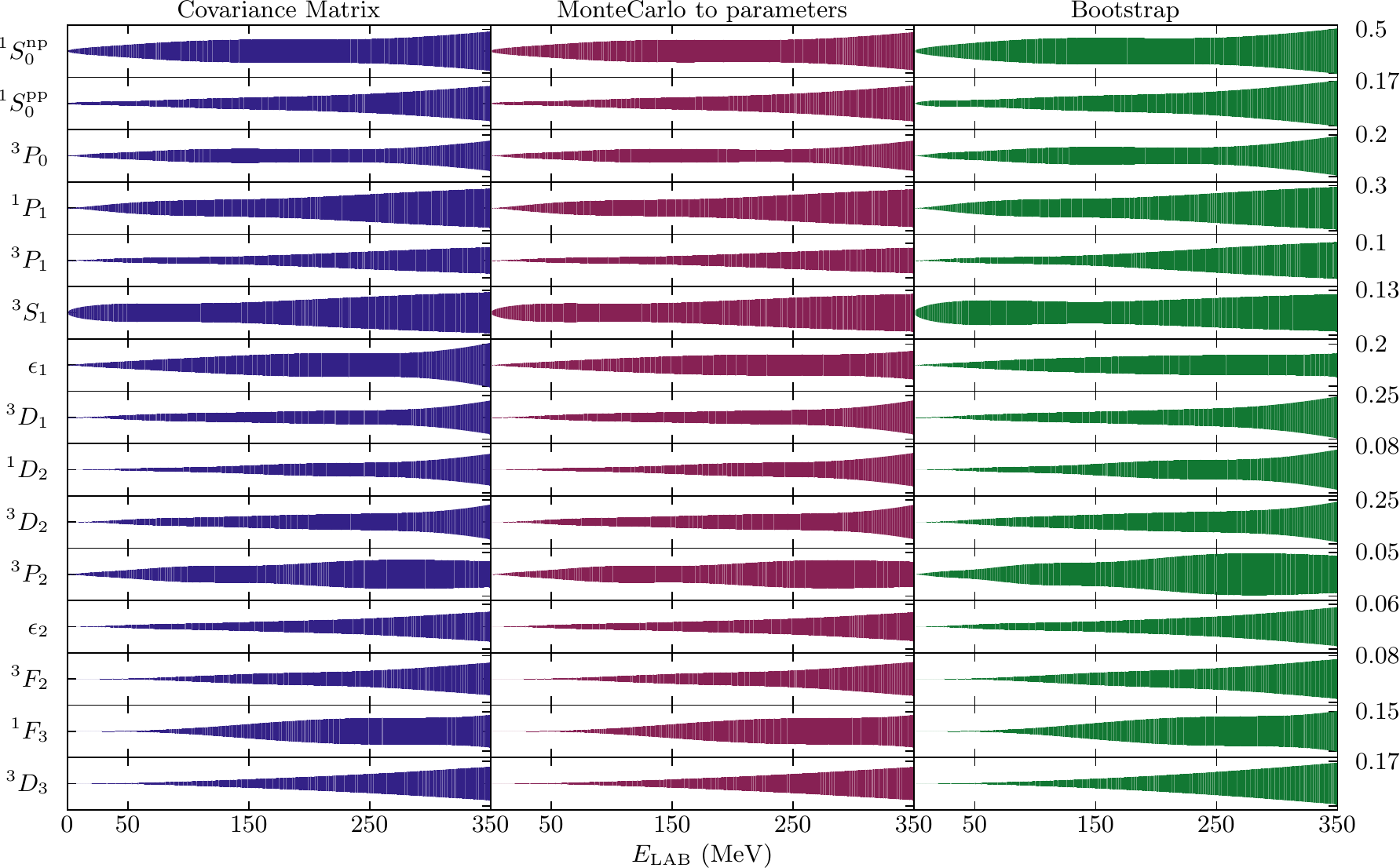}
\caption{(Color online) Low angular momentum partial wave phase-shifts
  statistical error bars calculated from the DS-OPE potential using
  the covariance matrix with Eq.(~\ref{eq:CovarianceErrors}) (left
  panel), a Monte Carlo sample of the potential parameters according
  to the same covariance matrix (middle panel) and the collection of
  parameters resulting from bootstrapping the database (right panel).}
\label{fig:PSerrorsbars}       
\end{figure*}

We compare in Fig.~\ref{fig:PSerrorsbars} the propagation of
statistical uncertainties into phase-shifts by three methods: i) the
standard covariance matrix method, ii) the equivalent Monte Carlo
implementation of the covariance matrix using the multivariate normal
distribution of Eq.~(\ref{eq:multinormal}) with $M=1020$ and iii) the
boostrap method also with $M=1020$ samples.  The first and the second
methods should produce the same results for a sufficietnly large
number of parameter samples. So the agreement between
Eq.(\ref{eq:CovarianceErrors}) and the Monte Carlo sampling of
parameters Eq.~(\ref{eq:multinormal}) reflects the large $M$ value
with the same ${\cal E}$. Although the bootstrap method tends to give
slightly larger error bars the difference with the other two methods
is not significant. As mentioned above, one potential advantage of the
Bootstrap method is that it relaxes the assumption of normally
distributed fitted parameters, a feature which proves relevant for
asymmetric or skewed distributions. We find that the asymmetries seen
in Figure \ref{fig:J1Distributions} do not significanly propagate to
the corresponding phase shifts. 

As a matter of principle the Monte Carlo simulation of data gives the
most reliable uncertainty propagation, but considering that performing
a large number of full-length fits to data can be computationally
expensive the covariance matrix methods are a fairly good and
extremely useful approximation which will be exploited in future work.

\section{Conclusions}

The propagation of statistical errors of nuclear forces stemming from
the finite precision and number of experimental NN scattering data
requires in the first place passing a normality test. However, even in
this favourable case the actual calculation may be computationally
demanding because of a practical need of repeating large scale
computations.  It is thus important to explore methods where the
number of calculations can be kept to a minimum. In the standard
covariance matrix method one needs the evaluation of the Hessian as
well as the derivatives of the object function whose uncertainties are
evaluated with respect to the theoretical model parameters. As an
alternative the Monte Carlo method based on explicit knowledge of the
Hessian can profitably be used as it avoids the computation of
derivatives (analytical or numerical) and automatically implements in
any snapshot the inherent correlations in the fitting parameters. The
previous methods assume a multivariate normal distribution of the
fitting parameters.

We have thus analyzed the more ellaborated bootstrap method which also
rests on the normality test and is based on a multiple minimization to
a synthetic set of data generated by the distribution of the most
likely estimate of the model parameters. While this approach assumes
normality of the experimental data but not of the fitting parameters,
it allows to handle possible skewness in the parameter
distributions. Our bootstrap analysis confirms the error and
correlations already found by the covariance method.

\vskip.1cm
This work is supported by Spanish DGI (grant FIS2011-24149) and Junta
de Andaluc{\'{\i}a} (grant FQM225).  R.N.P. is supported by a Mexican
CONACYT grant.


\end{document}